\documentclass[preprint,showpacks,preprintnumbers]{revtex4}
\usepackage{graphicx}
\usepackage{dcolumn}
\usepackage{bm}
\usepackage{amsmath, amsthm, amssymb}

\newtheorem{thm}{Theorem}[section]
\newtheorem{cor}[thm]{Collary}
\newtheorem{lem}[thm]{Lemma}
\newtheorem*{Crum}{Theorem 1 [Crum]}
\newtheorem*{Crum2}{Lemma 1 [Crum]}
\newtheorem*{Jacobi}{Theorem [Jacobi]}

\begin{document}

\title{Shape Invariance Through Crum Transformation}
\author{Jos{\'e} Orlando Organista, Marek Nowakowski,}
\affiliation{Departamento de Fisica, Universidad de los Andes, Cra.1E No. 18A-10, Santaf%
\'{e} de Bogot\'{a}, Colombia}
\author{H.C. Rosu}
\affiliation{Potosinian Institute of Science and Technology, Apartado Postal 3-74
Tangamanga, 78231 San Luis Potos\'{\i}, Mexico\\
}
\date{\today }

\begin{abstract}
We show in a rigorous way that Crum's result regarding the equal eigenvalue spectrum of Sturm-Liouville
problems can be obtained iteratively by successive Darboux transformations.
Furthermore, it can be shown that all neighbouring Darboux-transformed potentials of
higher order, $u_{k}$ and $u_{k+1}$, satisfy the condition of shape
invariance provided the original potential $u$ does so. Based on this result, we
prove that under the condition of shape invariance, the $n^{\rm th}$ iteration of
the original Sturm-Liouville problem defined solely through the shape invariance is
equal to the $n^{\rm th}$ Crum transformation.\newline
\newline
\vspace*{10pt}
%{\bf PACS} numbers: Mathematical Physics.%87.10.+e, 05.45.-a
\bigskip

\bigskip \bigskip \bigskip %{\flushright $PottsPT_{Nov05}$}
\end{abstract}

\maketitle

%................................................
%\begin{figure}[htb]
%\centerline{\includegraphics[scale=0.2]{openwebmail.eps}}
%\caption{IPICyT}
%\end{figure}
%....................................................


%\centerline{ArXiv: physics/051000x [v1]}

%\centerline{Physics Today: November 2005, pp. 19-21}

%,,,,,,,,,,,,,,,,,,,,,,,,,,,,,,,,,,,,,,,,,,,,,,,,,,,,,,,,,,,,,,,,,,,,,,,,,,,,,

%,,,,,,,,,,,,,,,,,,,,,,,,,,,,,,,,,,,,,,,,,,,,,,,,,,,,,,,,,,,,,,,,,,,,,,,,,,,,,,

\vspace*{10pt} %\keywords{The contents of the keywords}

%%111111111111111111111111111111111111111111111111111111111111111111111111
%%%%%%%%%%%%%%%%%%%%%%%%%%%%%%%%%%%
%%%%%%%%%%%%%%%%%%%%%%%%%%%%%%

\section{Introduction}
Supersymmetric quantum mechanics \cite{Nicolai, Witten}, the
factorization method \cite{Infeld}, the Darboux transformation
\cite{Darboux}, Crum's generalization of the former results
\cite{Crum}, the isospectral Hamiltonians based on the
Gelfand-Levitan equation \cite{GM, AM, Deift1, Pursey1} or the
Marchenko equation \cite{M, Pursey2} and the shape invariance
condition on the potentials \cite{Gendenshtein} together with a
transformation defined through this condition have been in the last
two decade an active area of mathematical physics \cite{Barklay,
Sukhatme, Carinena, Faux, Sasaki} and pure mathematics \cite{Deift2,
Gesztesy1, Gesztesy2}. The main concern of these areas has been the
construction of isospectral Schr\"odinger operators and the
analytical solvability of the Sturm-Liouville problem. The field
allowed a deeper insight into the eigenvalue problem and served as a
source for many new ideas and generalizations \cite{Aleixo,
Combescure, Rudyak, Schnitzer}. Indeed, it is almost impossible to
quote all research papers on the subject (suffices to note that one
review \cite{Cooper} and several books have been devoted to the
subject \cite{Junker, Bagchi, Matveev, Cooper2001, Cycon}). The
applications range from constructing new solvable potentials in
quantum
mechanics, differential equations \cite{Ince}, atomic physics \cite{Debergh}%
, nuclear physics \cite{Iachello}, classical mechanics \cite{Nowakowski},
acoustic spectral problems \cite{Adler} to
quantum gravitation \cite{graham, socorro} and neutrino oscillation \cite%
{Balantekin}, to mention a few important areas.

Mathematically, not all these transformations mentioned above are equal, or at
least this is not apparent at first sight. For instance, the usual
Darboux transformation is not the most general solution of the Riccati
equation and as such does not give us the most general transformation in
connection with the isospectral eigenvalue spectrum. On the other hand, the
generalization of the Darboux transformations, namely, the so-called Crum
transformation appears to be much more complicated than the original Darboux
result and as such seems to offer us new avenues to construct new
potentials. The third transformation of a Hamiltonian which we have in mind (defined here in
equation (92)) is closely related to the condition of shape invariance.
Hence, without doubt, there is some need to at least classify these
transformations according to the complexity or generality and to uncover
their relations between them. One such result in this direction is the
nonequivalence of the Abraham-Moses \cite{AM} and Darboux constructions shown
in \cite{Pursey1}.
Two remarks are in order here. Firstly, it is understood that unlike the Darboux
transformation, any transformation in connection with the shape invariance
is, of course, limited to the set of shape invariant Hamiltonians.
Secondly, for completeness it is worth noting that the level of complexity of isospectral quantum systems
can be increased by considering  non-linear and higher order
supersymmetric transformations \cite{A1, A2, A3, A4}. These are transformations which
cannot be reached by iterative Darboux transformations.
In this work, however,  we will not consider these kind of
transformations and restrict ourselves to the Darboux case.
After some preparatory statements we will show that the
undertaking to uncover relationships between the transformations gives a
simple result, namely, allowing the use of higher order Darboux
transformations, we can state that all three transforms of the original
Sturm-Liouville problem are equal. This result is based on a theorem which
we prove in the present paper concerning higher order Darboux
transformations of shape invariant potentials denoted by $u^{D}[k]$. The
theorem states that provided the original potential satisfies the shape
invariance conditions, all pairs $u^{D}[k]$, $u^{D}[k+1]$ are also mutually
shape invariant. The theorem can be proved by induction. Interestingly, it
intertwines this induction with another statement, this time for the wave
functions.
We illustrate the theorems by two examples.

%.......................2222222222222222222222222222222222222222222222222222222222222

\section{Crum's Result}

In this section, we briefly present Crum's result and comment on one
identity on which Crum's result is partly based. This identity is crucial
for the subsequent results which we will elaborate upon in the next section.

Let
\begin{equation}
W_{k}\equiv W\left( \psi _{1},\psi _{2},...,\psi _{k}\right) =\det A\quad
,\quad A_{ij}=\frac{d^{i-1}\psi _{j}}{dx^{i-1}}\quad i,j=1,2,...,k ~,
\label{110}
\end{equation}
be the Wronskian determinant of the functions $\psi _{1},\psi _{2},...,\psi _{k}$, and
\begin{equation}
W_{k,s}=W\left( \psi _{1},\psi _{2},...,\psi _{k},\psi _{s}\right) ~.
\end{equation}

%\begin{thm} \label{Crum}

\begin{Crum}
If $\psi _{1},\psi _{2},...,\psi _{n}$ are the solutions of the regular
Sturm-Liouville problem
\begin{equation}
-\frac{d^{2}\psi _{s}}{dx^{2}}+u\psi _{s}=\lambda _{s}\psi _{s}  \label{109}
\end{equation}%
then $\psi^C \left[ n\right] _{s}$ satisfies the Sturm-Liouville equation
\begin{equation}
-\frac{d^{2}\psi ^{C}[n]_{s}}{dx^{2}}+u^{C}[n]\,\psi ^{C}[n]_{s}=\lambda
_{s}\psi ^{C}[n]_{s}~,  \label{113}
\end{equation}%
with $\psi ^{C}[n]_{s}$ and $u^{C}[n]_{s}$ given by
\begin{equation}
\psi _{s}\longrightarrow \psi ^{C}[n]_{s}\equiv \frac{W_{n,s}}{W_{n}}
\label{64}
\end{equation}%
and
\begin{equation}
u\longrightarrow u^{C}[n]=u-2\frac{d^{2}}{dx^{2}}\ln W_{n} ~.
\end{equation}
\end{Crum}

Note that the Crum transforms of $\psi $ and $u$ are not defined
iteratively. By $[C]$ we wish to distinguish the Crum transformation from
other transforms (like Darboux) which will be defined later in the text. The
proof of Crum's theorem can be found in \cite{Crum} and \cite{Matveev}. We
comment here only on one cornerstone of the original proof given by Crum
\cite{Crum} which we will also use later. The first step in the proof of
Crum's result on the Wronskian determinant is to consider the derivative of $%
W_{k}$. Taking the derivative of
\begin{equation}
W_{2}=\left\vert
\begin{array}{cc}
\psi _{1} & \psi _{2} \\
\frac{d\psi _{1}}{dx} & \frac{d\psi _{2}}{dx}%
\end{array}%
\right\vert ~,
\end{equation}%
we find the rather obvious result,
\begin{equation}
\frac{dW_{2}}{dx}=\left\vert
\begin{array}{cc}
\psi _{1} & \psi _{2} \\
\psi _{1}^{\prime \prime } & \psi _{2}^{\prime \prime }%
\end{array}%
\right\vert ~,
\end{equation}%
where we used the notation $\psi ^{\prime \prime }_i$ for $\frac{d^{2}\psi _{i}%
}{dx^{2}}.$, $i=1,2$. This result can be readily generalized for the $n\times n$ case.

\begin{lem}
\label{wronsky} For the derivative of a Wronskian determinant we have
\begin{equation}
W\,_{n}^{\prime }=\left\{ \psi _{1}^{\left( n\right) }M_{\left( 1,n\right)
}^{\left( 1\right) }+\psi _{2}^{\left( n\right) }M_{\left( 2,n\right)
}^{\left( 1\right) }+\cdots +\psi _{n-1}^{\left( n\right) }M_{\left(
n-1,n\right) }^{\left( 1\right) }+\psi _{n}^{\left( n\right) }M_{\left(
n,n\right) }^{\left( 1\right) }\right\}~.  \label{199}
\end{equation}
\end{lem}

Assume the result to be valid for $n-1$. Using the Laplace expansion
according to the last line of the Wronskian $W_{n}$ we get
\begin{equation*}
W\,_{n}^{\prime }=
\end{equation*}
\begin{eqnarray}  \label{derivative}
&&\left\{ \psi _{1}^{\left( n\right) }M_{\left( 1,n\right) }^{\left(
1\right) }+\psi _{2}^{\left( n\right) }M_{\left( 2,n\right) }^{\left(
1\right) }+\cdots +\psi _{n-1}^{\left( n\right) }M_{\left( n-1,n\right)
}^{\left( 1\right) }+\psi _{n}^{\left( n\right) }M_{\left( n,n\right)
}^{\left( 1\right) }\right\} +\bigskip \\
&&\left\{ \psi _{1}^{\left( n-1\right) }\left( M_{\left( 1,n\right)
}^{\left( 1\right) }\right) ^{\prime }+\cdots +\psi _{n-1}^{\left(
n-1\right) }\left( M_{\left( n-1,n\right) }^{\left( 1\right) }\right)
^{\prime }+\psi _{n}^{\left( n-1\right) }\left( M_{\left( n,n\right)
}^{\left( 1\right) }\right) ^{\prime }\right\}  \notag
\end{eqnarray}
where every $(n-1) \times (n-1)$ determinant $M_{\left(i,
n\right)}^{\left(1\right)}$ is a Wronskian for which, by assumption, the
theorem is valid. Hence
\begin{equation}
\mathrm{det}B\equiv \left\{ \psi _{1}^{\left( n-1\right) }\left( M_{\left(
1,n\right) }^{\left( 1\right) }\right) ^{\prime }+\cdots +\psi
_{n-1}^{\left( n-1\right) }\left( M_{\left( n-1,n\right) }^{\left( 1\right)
}\right) ^{\prime }+\psi _{n}^{\left( n-1\right) }\left( M_{\left(
n,n\right) }^{\left( 1\right) }\right) ^{\prime }\right\}
\end{equation}
is a determinant whose two last lines are equal and therefore $\mathrm{det}%
B=0$. The result (\ref{derivative}) can be written as
\begin{equation}
W\,_{n}^{\prime }=\left|
\begin{array}{ccccc}
\psi _{1} & \psi _{2} & \cdots & \psi _{n-1} & \psi _{n}\medskip \\
\psi _{1}^{\prime } & \psi _{2}^{\prime } & \cdots & \psi _{n-1}^{\prime } &
\psi _{n}^{\prime }\medskip \\
\vdots & \vdots & \cdots & \vdots & \vdots \medskip \\
\psi _{1}^{\left( n-2\right) } & \psi _{2}^{\left( n-2\right) } & \cdots &
\psi _{n-1}^{\left( n-2\right) } & \psi _{n}^{\left( n-2\right) }\medskip \\
\psi _{1}^{\left( n\right) } & \psi _{2}^{\left( n\right) } & \cdots & \psi
_{n-1}^{\left( n\right) } & \psi _{n}^{\left( n\right) } %
\end{array}
\right|~.  \label{192}
\end{equation}

We can now state a result which will be of some importance later and which
is one of the important ingredients in proving Theorem 1 of Crum.

\begin{Crum2}
The Wronski determinant of the two Wronskians, $W_{n}$ and $W_{n-1,s}$, is
equal to $W_{n,s}W_{n-1}$. In other words
\begin{equation}
W\left( W_{n},W_{n-1,s}\right) =W_{ns}W_{n-1}~.  \label{82}
\end{equation}
\end{Crum2}

The proof relies on the Jacobi theorem for determinants (see Appendix A). We
refer the reader to Appendix A for the proof of this Lemma too.

It is well known that for $n=1$ the Crum transformations reduce to the Darboux
transformation when $W_{1}=\psi _{1},$ and $W_{1,s}=W\left( \psi _{1},\psi
_{s}\right) $. Specifically, we have
\begin{equation}
\psi ^{D}[1]_{s}\equiv \psi ^{C}[1]_{s}=\frac{W_{1,s}}{\psi _{1}}=\psi
_{s}^{\prime }-\frac{\psi _{1}^{\prime }}{\psi _{1}}\psi _{s}\quad s>1 ~,
\label{crumdar1}
\end{equation}%
\begin{equation}
u^{D}[1]\equiv u^{C}\left[ 1\right] =u-2\frac{d^{2}}{dx^{2}}\ln W_{1}=u-2%
\frac{d}{dx}\frac{\psi _{1}^{\prime }}{\psi _{1}}~.  \label{crumdar2}
\end{equation}%
We can define higher order Darboux transformations iteratively by
\begin{equation}
\left\{
\begin{array}{c}
u^{D}\left[ k-1\right] \rightarrow u^{D}\left[ k\right] =u\left[ k-1\right]
-2\dfrac{d}{dx}\dfrac{\left( \psi ^{D}\left[ k-1\right] _{k}\right) ^{\prime
}}{\psi ^{D}\left[ k-1\right] _{k}}~,\bigskip  \\
\psi ^{D}\left[ k-1\right] _{s}\rightarrow \psi ^{D}\left[ k\right] _{s}=%
\dfrac{W\left( \psi ^{D}\left[ k-1\right] _{k}\,,\,\psi ^{D}\left[ k-1\right]
_{s}\right) }{\psi ^{D}\left[ k-1\right] _{k}}~.%
\end{array}%
\right. \quad s>k  \label{132}
\end{equation}%
Obviously, the last equation can be written also in a way which resembles
more the first Darboux transformation, i.e.,
\begin{equation}
\psi ^{D}\left[ k\right] _{s}=(\psi ^{D}\left[ k-1\right] _{s})^{\prime}-\frac{%
(\psi ^{D}\left[ k-1\right] _{k})^{\prime}}{\psi ^{D}\left[ k-1\right] _{k}}\psi
^{D}\left[ k-1\right] _{s}~.
\end{equation}

It is a priori not clear as to what connection the $k$-th Darboux transformation
has with the $k$-th Crum transformation and if they can be related at all,
except for the definition at the lowest order of Crum's transformation. The
answer is provided in the next section.

%.........................................................................33333333333333333333333333333333333

\section{The connection between higher order Darboux and Crum transformation}
To this end,
let us first examine the simplest case of $k=2$:%
\begin{equation}
u^{D}\left[ 2\right] =u^{D}\left[ 1\right] -2\frac{d}{dx}\frac{\left( \psi
^{D}\left[ 1\right] _{2}\right) ^{\prime }}{\psi ^{D}\left[ 1\right] _{2}}~.
\label{117}
\end{equation}%
Since $u[D]_{\left[ 1\right] }$ is the Darboux transformed potential the above equation reads
\begin{equation}
u^{D}\left[ 2\right] =u-2\frac{d}{dx}\left( \frac{\psi _{1}^{\prime }}{\psi
_{1}}+\frac{(\psi ^{D}\left[ 1\right] _{2})^{\prime}}{\psi ^{D}\left[ 1\right]
_{2}}\right)~.   \label{179}
\end{equation}%
According to (19), $\psi ^{D}\left[ 1\right] _{2}=\frac{W_{1,2}}{\psi _{1}}$
and on account of the simple identity $W_{n,n+1}=W_{n+1}$, we can write,
\begin{equation}
u^{D}\left[ 2\right] =u-2\frac{d}{dx}\left( \frac{\psi _{1}^{\prime }}{\psi
_{1}}+\frac{\left( \frac{W_{2}}{\psi _{1}}\right) ^{\prime }}{\frac{W_{2}}{%
\psi _{1}}}\right) ~,
\end{equation}%
which finally gives
\begin{equation}
u^{D}\left[ 2\right] =u-2\frac{d}{dx}\left( \frac{W_{2}^{\,\prime }}{W_{2}}%
\right) =u^{C}\left[ 2\right]~.
\end{equation}%
Similarly, the eigenfunctions
\begin{equation}
\psi ^{D}\left[ 2\right] _{s}=\frac{\left\vert
\begin{array}{cc}
\psi ^{D}\left[ 1\right] _{2}\medskip  & \psi ^{D}\left[ 1\right]
_{s}\medskip  \\
\frac{d}{dx}\medskip \psi ^{D}\left[ 1\right] _{2} & \frac{d}{dx}\psi ^{D}%
\left[ 1\right] _{s}%
\end{array}%
\right\vert }{\psi ^{D}\left[ 1\right] _{2}}
\end{equation}%
can be cast into the form
\begin{equation}
\psi ^{D}[2]_{s}=\frac{\frac{1}{\psi _{1}}\left\vert
\begin{array}{cc}
W_{2}\medskip  & W_{1,s} \\
\left( \frac{W_{2}}{\psi _{1}}\right) ^{\prime } & \left( \frac{W_{1,s}}{%
\psi _{1}}\right) ^{\prime }%
\end{array}%
\right\vert }{\frac{W_{2}}{\psi _{1}}}=\frac{\left\vert
\begin{array}{cc}
W_{2}\medskip  & W_{1,s} \\
\left( \frac{W_{2}}{\psi _{1}}\right) ^{\prime } & \left( \frac{W_{1,s}}{%
\psi _{1}}\right) ^{\prime }%
\end{array}%
\right\vert }{W_{2}}~.
\end{equation}%
With the help of the standard property of determinants, namely $\mathrm{det}%
\left( \vec{z}_{1},...,\vec{z}_{i},...,\vec{z}_{n}\right) =\mathrm{det}%
\left( \vec{z}_{1},...,\vec{z}_{i}+\alpha \vec{z}_{k},...,\vec{z}_{n}\right)
$ the last equation reduces to
\begin{equation}
\psi ^{D}\left[ 2\right] _{s}=\frac{\frac{1}{\psi _{1}}\left\vert
\begin{array}{ll}
W_{2}\medskip  & W_{1,s} \\
W_{2}^{\,\prime } & W_{1,s}^{\,\prime }%
\end{array}%
\right\vert }{W_{2}}~.
\end{equation}

Applying the result of Lemma 1 and remembering that $W_{1}=\psi _{1}$, one finally finds
\begin{equation}
\psi ^{D}\left[ 2\right] _{s}=\frac{W_{2,s}}{W_{2}}=\psi ^{C}\left[ 2\right]
_{s}~.
\end{equation}%
The steps above will serve as a beginning of the induction proof of the
following general statement:

\begin{thm}
The n-th Crum transformation is equivalent to the n-th higher order Darboux
transformation. This is to say, any Crum transformation can be reached
iteratively by successive Darboux transformations, i.e.,
\end{thm}

\begin{eqnarray}
u^{C}\left[ n\right]  &=&u^{D}\left[ n\right]~,    \notag \\
\psi ^{C}\left[ n\right] _{s} &=&\psi ^{D}\left[ n\right] _{s}~.
\end{eqnarray}%
\emph{Proof}. Assuming the theorem to be valid for $n$ means that the
statement
\begin{equation}
\left\{
\begin{array}{c}
u^{D}\left[ n\right] =u^{D}\left[ n-1\right] -2\dfrac{d}{dx}\dfrac{(\psi
^{D}\left[ n-1\right] _{n})^{\prime}}{\psi ^{D}\left[ n-1\right] _{n}}\bigskip
\\
\psi ^{D}\left[ n\right] _{s}=\dfrac{W\left( \psi ^{D}\left[ n-1\right]
_{n}\,,\psi ^{D}\left[ n-1\right] _{s}\right) }{\psi ^{D}\left[ n-1\right]
_{n}}%
\end{array}%
\right. \quad s>n
\end{equation}%
is equivalent to
\begin{equation}
\left\{
\begin{array}{c}
u^{D}\left[ n\right] =u-2\dfrac{d}{dx}\left( \dfrac{W_{n}^{\,\prime }}{W_{n}}%
\right) \bigskip  \\
\psi ^{D}\left[ n\right] _{s}=\dfrac{W_{n,s}}{W_{n}}~.%
\end{array}%
\right. \quad s>n
\end{equation}%
Based on that, we have to show
\begin{equation}
u^{C}\left[ n+1\right] =u^{D}\left[ n+1\right] =u^{D}\left[ n\right] -2%
\dfrac{d}{dx}\dfrac{(\psi ^{D}\left[ n\right] _{n+1})^{\prime}}{\psi ^{D}\left[ n%
\right] _{n+1}}=u-2\frac{d}{dx}\frac{W_{n+1}^{\prime }}{W_{n+1}}~,   \label{137}
\end{equation}%
and
\begin{equation}
\psi ^{D}\left[ n+1\right] _{s}=\dfrac{W\left( \psi ^{D}\left[ n\right]
_{n+1},\psi ^{D}\left[ n\right] _{s}\right) }{\psi ^{D}\left[ n\right] _{n+1}%
}=\frac{W_{n+1,s}}{W_{n+1}}~.
\end{equation}%
The validity of the hypothesis of the induction for $n$ allows us to write
\begin{equation}
u^{D}\left[ n+1\right] =u^{D}\left[ n\right] -2\frac{d}{dx}\frac{(\psi
^{D}\left[ n\right] _{n+1})^{\prime}}{\psi ^{D}\left[ n\right] _{n+1}}=u-2\frac{%
d}{dx}\left( \frac{W_{n}^{\,\prime }}{W_{n}}+\frac{(\psi ^{D}\left[ n%
\right] _{n+1})^{\prime}}{\psi ^{D}\left[ n\right] _{n+1}}\right)~.
\end{equation}%
Using the validity of the hypothesis for $n$, but this time for the wave
functions, implies
\begin{equation}
u^{D}\left[ n+1\right] =u-2\frac{d}{dx}\left( \frac{W_{n}^{\,\prime }}{W_{n}}%
+\frac{\left( \frac{W_{n+1}}{W_{n}}\right) ^{\prime }}{\frac{W_{n+1}}{W_{n}}}%
\right) =u-2\frac{d}{dx}\left( \frac{W_{n+1}^{\,\prime }}{W_{n+1}}\right)
=u^{C}\left[ n+1\right]~.
\end{equation}%
Similarly, the result for the eigenfunctions may be written as $\bigskip $%
\begin{equation}
\psi ^{D}\left[ n+1\right] _{s}=\frac{\left\vert
\begin{array}{ccc}
\psi ^{D}\left[ n\right] _{n+1} &  & \psi ^{D}\left[ n\right] _{s} \\
&  &  \\
\psi ^{\prime D}\left[ n\right] _{n+1} &  & \psi ^{\prime D}\left[ n\right]
_{s}%
\end{array}%
\right\vert }{\psi ^{D}\left[ n\right] _{n+1}}=\frac{\left\vert
\begin{array}{ccc}
\frac{W_{n+1}}{W_{n}} &  & \frac{W_{n,s}}{W_{n}} \\
&  &  \\
\left( \frac{W_{n+1}}{W_{n}}\right) ^{\prime } &  & \left( \frac{W_{n,s}}{%
W_{n}}\right) ^{\prime }%
\end{array}%
\right\vert }{\frac{W_{n+1}}{W_{n}}}~.
\end{equation}%
$\bigskip $ One easily proceeds now to verify the validity of the following equation
\begin{equation}
\psi^D \left[ n+1\right]_s=\frac{\left\vert
\begin{array}{ccc}
W_{n+1} &  & W_{n,s} \\
&  &  \\
\frac{W_{n+1}^{\,\prime }}{W_{n}}-\frac{W_{n}^{\,\prime }W_{n+1}}{W_{n}^{2}}
&  & \frac{W_{n,s}^{\,\prime }}{W_{n}}-\frac{W_{n}^{\,\prime }W_{n,s}}{%
W_{n}^{2}}%
\end{array}%
\right\vert }{W_{n+1}}=\frac{\frac{1}{W_{n}}\left\vert
\begin{array}{ccc}
W_{n+1} &  & W_{n,s} \\
&  &  \\
W_{n+1}^{\,\prime } &  & W_{n,s}^{\,\prime }%
\end{array}%
\right\vert }{W_{n+1}}~.
\end{equation}%
By virtue of Lemma 1 we can assure that
\begin{equation}
\psi ^D\left[ n+1\right]_s=\frac{W_{n+1,s}}{W_{n+1}}=\psi^C \left[ n+1\right]_s
\end{equation}%
is true which completes the proof.

It is instructive to follow this theorem by an explicit examples.

%.....................4444444444444444444444444444444444444444444444444444444444444444444444

\section{Two Examples}
In this section we will demonstrate the above theorems by two examples. We choose first
a potential which
satisfies the condition of shape invariance (Morse potential) followed by a simple example
which falls into the class
of non-shape invariant, but solvable potentials.

Let us consider, as an example, the Sturm Liouville problem with the Morse
potential, i.e.,
\begin{equation}
u\left( x;A\right) =2\left[ A^{2}-A\left( A+\frac{\alpha }{\sqrt{2}}\right)
\mathrm{sech}{}^{2}\left( \alpha x\right) \right]~.  \label{mor}
\end{equation}%
The super-partner of this potential corresponding to $\psi_1$ and $\lambda_1$ is
\begin{equation} \label{extra1}
u^{C}\left[ 1\right] =u^{D}\left[ 1\right] =2\left[ A^{2}-AA_{1}\mathrm{sech}%
{}^{2}\left( \alpha x\right) \right]
\end{equation}%
and the first three eigenfunctions are given by

\begin{enumerate}
\item $\psi _{1}=c_{1}\left[ \mathrm{sech}\left( \alpha x\right) \right] ^{%
\frac{\sqrt{2}A}{\alpha }}~,\bigskip $

\item $\psi _{2}=c_{2}\sinh \left( \alpha x\right) \psi _{1}~,\bigskip $

\item $\psi _{3}=c_{3}\left( -\cosh ^{2}\left( \alpha x\right) +\frac{\left(
2\sqrt{2}A-\alpha \right) }{\alpha }\sinh ^{2}\left( \alpha x\right) \right)
\psi _{1}~.$
\end{enumerate}
In the following we will not determine the constants $c_i$ as they are
of minor importance for our results. Secondly, the results become increasingly complicated.
For instance, to calculate $c_1$ we can use
\begin{equation} \label{c1}
\int_0^{\infty} {\rm sech}(ax)^{\frac{2\sqrt{2}A}{\alpha}}dx=-\frac{1}{\sqrt{2}A}\, { }_2F_1
\left(\frac{\sqrt{2}A}{\alpha}, \frac{1}{2},
1+\frac{\sqrt{2}A}{\alpha}, \frac{\alpha}{\sqrt{2}A}\right)
\end{equation}
where $_2F_1$ is the hypergeometric function.
The corresponding eigenvalues can be compactly written as
\begin{equation}
\lambda _{n}=2\left( A^{2}-\left( A-\frac{\left( n-1\right) \alpha }{\sqrt{2}%
}\right) ^{2}\right)~. \quad
\end{equation}%
It is convenient to define $A_{n}$ as
\begin{equation} \label{extray}
A_{n}\equiv A-\frac{n\alpha }{\sqrt{2}}
\end{equation}%
such that the eigenvalues read now
\begin{equation} \label{eigenvaluesx}
\lambda _{n}=2\left( A^{2}-A_{n-1}^{2}\right) \quad n=1,2,3,... \,\,\, ~.
\end{equation}%
The first three are explicitly given as follows
\begin{equation}
\lambda _{1}=0\quad ,\quad \lambda _{2}=2\sqrt{2}A\alpha -\alpha ^{2}\quad
,\quad \lambda _{3}=4\sqrt{2}A\alpha -4\alpha ^{2}~.  \label{vpropios}
\end{equation}%
Besides equation (\ref{extra1}) we will also need the following results:
\begin{eqnarray}
%u^{C}\left[ 1\right]  &=&u^{D}\left[ 1\right] =2\left[ A^{2}-AA_{1}\mathrm{%
%sech}{}^{2}\left( \alpha x\right) \right]   \notag  \label{psi13} \\
\psi ^{C}\left[ 1\right] _{2} &=&\psi ^{D}\left[ 1\right] _{2}=c_{2}\alpha
\cosh \left( \alpha x\right) \psi _{1},   \label{extra42}\\
\label{extra43}
\psi ^{C}\left[ 1\right] _{3}=\psi ^{D}\left[ 1\right] _{3} &=&4\sqrt{2}%
c_{3}A_{1}\sinh \left( \alpha x\right) \cosh \left( \alpha x\right) \psi
_{1}~.
\end{eqnarray}%
To show explicitly the equality $u^{C}\left[ 2\right] =u^{D}\left[ 2\right] $,
we start with $u^{C}\left[ 2\right] $, i.e.,
\begin{equation}
u^{C}\left[ 2\right] =u-2\frac{d^{2}}{dx^{2}}\ln W_{2}~.
\end{equation}%
Since $W_{1,2}=W_{2}$ and using
\begin{equation}
\frac{W_{1,2}}{W_{1}}=c_{2}\alpha \cosh \left( \alpha x\right) \psi _{1}
\label{w12}
\end{equation}%
we have,
\begin{equation}
\ln W_{2}=\ln c_{2}\alpha +\ln \cosh \left( \alpha x\right) +2\ln \psi _{1}~,
\end{equation}%
but also
\begin{equation}
\frac{d^{2}}{dx^{2}}\ln W_{2}=\alpha \left( \alpha -2\sqrt{2}A\right)
\mathrm{sech}{}^{2}\left( \alpha x\right)~.
\end{equation}%
Finally, with (\ref{mor}) we arrive at
\begin{equation}
u^{C}\left[ 2\right] =2\left[ A^{2}-A_{1}A_{2}\mathrm{sech}{}^{2}\left(
\alpha x\right) \right]~.
\end{equation}%
Next we turn to the expression for $u^D \left[ 2\right] $, namely,
\begin{equation} \label{extra100}
u^{D}\left[ 2\right] =u^{D}\left[ 1\right] -2\frac{d}{dx}\frac{(\psi ^{
D}\left[ 1\right] _{2})^{\prime}}{\psi ^{D}\left[ 1\right] _{2}}~.
\end{equation}%
Taking into account equation (\ref{extra42}) we obtain
\begin{equation}
\frac{(\psi ^{D}\left[ 1\right] _{2})^{\prime}}{\psi ^{D}\left[ 1\right] _{2}}=-%
\sqrt{2}A_{1}\tanh \left( \alpha x\right)
\end{equation}%
and therefore
\begin{equation}
\frac{d}{dx}\frac{(\psi ^{D}\left[ 1\right] _{2})^{\prime}}{\psi ^{D}\left[ 1%
\right] _{2}}=-\sqrt{2}\alpha A_{1}\mathrm{sech}{}^{2}\left( \alpha x\right)~.
\end{equation}%
From this we conclude (see (\ref{extra100})) that
\begin{equation}
u^{D}\left[ 2\right] =2\left[ A^{2}-A_{1}A_{2}\mathrm{sech}{}^{2}\left(
\alpha x\right) \right]
\end{equation}%
and hence
\begin{equation}
u^{C}\left[ 2\right] =u^{D}\left[ 2\right]~.   \label{udarcrum2}
\end{equation}%
The transformed potentials here have almost identical functional form. This
is, of course, due to the choice of the potential and need not be so in
other cases.

To demonstrate that $\psi ^{C}\left[ 2\right] _{s}=\psi ^{D}\left[ 2\right]
_{s}$, we calculate $\psi ^{C}\left[ 2\right] _{3}$ to be
\begin{equation}
\psi ^{C}\left[ 2\right] _{3}=\frac{W_{3}}{W_{2}}=\frac{\left\vert
\begin{array}{ccc}
\psi _{1} & \psi _{2} & \psi _{3} \\
\psi _{1}^{\prime } & \psi _{2}^{\prime } & \psi _{3}^{\prime } \\
-\lambda _{1}\psi _{1} & -\lambda _{2}\psi _{2} & -\lambda _{3}\psi _{3}%
\end{array}%
\right\vert }{\left\vert
\begin{array}{cc}
\psi _{1} & \psi _{2} \\
\psi _{1}^{\prime } & \psi _{2}^{\prime }%
\end{array}%
\right\vert }=\lambda _{2}\psi _{2}\frac{W_{1,3}}{W_{2}}-\lambda _{3}\psi
_{3}~.
\end{equation}%
Making use of
\begin{equation}
\lambda _{2}\psi _{2}\frac{W_{1,3}}{W_{2}}=\left( \frac{4\sqrt{2}\lambda
_{2}c_{3}A_{1}}{\alpha }\right) \sinh ^{2}\left( \alpha x\right) \psi _{1}
\end{equation}%
this becomes
\begin{equation} \label{last2}
\psi ^{C}\left[ 2\right] _{3}=\lambda _{3}c_{3}\cosh ^{2}\left( \alpha
x\right) \psi _{1}\left( x\right)~.
\end{equation}
On the other hand
\begin{eqnarray} \label{definition}
\psi ^{D}\left[ 2\right] _{3} &=&\frac{
\left\vert
\begin{array}{cc}
\psi \left[ 1\right] _2 & \psi \left[ 1\right]_3 \\
(\psi \left[ 1\right]_2)^{\prime } & (\psi \left[ 1\right]_3)^{\prime }%
\end{array}%
\right\vert }
{\psi \left[ 1\right]_2}=
\frac{\left( \frac{\psi \left[ 1
\right]_3}{\psi \left[ 1\right]_2}\right) ^{\prime }(\psi \left[ 1
\right]_2)^{2}}{\psi \left[ 1\right]_{2}}=\left( \frac{\psi \left[ 1
\right]_3}{\psi \left[ 1\right]_2}\right) ^{\prime }\psi \left[ 1
\right]_2  \notag \\
&=&\lambda _{3}c_{3}\cosh ^{2}\left( \alpha x\right) \psi _{1}  \label{psi23}
\end{eqnarray}
where on the right hand side we already dropped the distinction between $D$ and $C$
(see (\ref{extra42}) and (\ref{extra43})).
The simple conclusion that we can draw is
\begin{equation}
\psi ^{C}\left[ 2\right] _{3}=\psi ^{D}\left[ 2\right] _{3}~.
\end{equation}

It is instructive to consider also a case of a solvable, but non-shape invariant potential.
Many such cases are known (see \cite{Natanzon, Ginocchio, Sing} and the discussion in \cite{Cooper})
and explicit proofs that these potentials fail to satisfy the
shape invariance condition were given. For instance, for the case of the Natanzon potential this was shown in
\cite{Cooper2}. Many of these potentials are complicated and some, like the Natanzon case, only known in implicit form.
Therefore, for the sake of efficient calculations,
it is recommendable to develop first a fast algorithm to perform the
desired calculations. We will do exactly that before explicitly giving the explicit example of the
Ginocchio case. Imagine we would like to calculate $\psi^D\left[2 \right]_3$.
In turns out that the calculation can be greatly simplified by
invoking the ratios $h_n=\psi_n^{\prime}/\psi_n$ where $\psi_n$ is, as usual, the eigenfunction to
the $\epsilon_n$ eigenvalue.
It is now a straightforward exercise to show that
\begin{equation} \label{G1}
\psi^D\left[ 2\right] _3=\frac{\left\vert
\begin{array}{cc}
\psi^D\left[ 1\right]_2\medskip  & \psi^D\left[ 1%
\right]_3 \\
(\psi^D\left[ 1\right]_2)^{\prime} & (\psi^D\left[ 1\right])_3)^{\prime}%
\end{array}%
\right\vert }{(\psi^D\left[ 1\right]_2)}
\end{equation}%
is equal to
\begin{eqnarray} \label{G2}
\frac{1}{(h_1-h_0)\psi_1}\biggl [
&&\left( h_{1}-h_{0}\right) \left\{ \left( h_{2}-h_{0}\right) ^{\prime
}+\left( h_{2}-h_{0}\right) h_{2}\right\} \psi _{1}\psi _{2}- \\
&&\left( h_{2}-h_{0}\right) \left\{ \left( h_{1}-h_{0}\right) ^{\prime
}+\left( h_{1}-h_{0}\right) h_{1}\right\} \psi _{1}\psi _{2} \biggr ] ~.
\end{eqnarray}%
Using the Schr\"odinger equation the latter simplifies to
\begin{equation} \label{G3}
\psi^D\left[ 2\right]_3=\frac{\left[ \epsilon _{0}\left\{
h_{2}-h_{1}\right\} -\epsilon _{1}\left\{ h_{2}-h_{0}\right\} +\epsilon
_{2}\left\{ h_{1}-h_{0}\right\} \right] }{\left( h_{1}-h_{0}\right) }\psi
_{2}~.
\end{equation}
It is obvious that, provided we know the functions $h_0, h_1, h_2$ and $\psi_2$, this expression
allows a fast calculation or the wave function $\psi^D[2]_3$ for arbitrary potential.
The Crum's result gives
\begin{equation} \label{G4}
\psi^C\left[ 2\right]_3=\frac{W_{2,3}}{W_{2}}=\frac{W_{3}}{W_{2}%
}=\frac{W_3}{(h_1-h_0)\psi_0 \psi_1}~,
\end{equation}%
where $W_3$ is
\begin{eqnarray} \label{G5}
W_3=\left\vert
\begin{array}{ccc}
\psi _{0}\medskip  & \psi _{1} & \psi _{2} \\
\psi _{0}^{\prime } & \psi _{1}^{\prime } & \psi _{2}^{\prime } \\
\psi _{0}^{\prime \prime } & \psi _{1}^{\prime \prime } & \psi _{2}^{\prime
\prime }%
\end{array}%
\right\vert
=
\left\vert
\begin{array}{ccc}
\psi _{0}\medskip  & \psi _{1} & \psi _{2} \\
h_{0}\psi _{0} & h_{1}\psi _{1} & h_{2}\psi _{2} \\
\psi _{0}^{\prime \prime } & \psi _{1}^{\prime \prime } & \psi _{2}^{\prime
\prime }%
\end{array}%
\right\vert
&=&
\left\vert
\begin{array}{ccc}
\psi _{0}\medskip  & \psi _{1} & \psi _{2} \\
h_{0}\psi _{0} & h_{1}\psi _{1} & h_{2}\psi _{2} \\
u-\epsilon _{0}\psi _{0} & u-\epsilon _{1}\psi _{1} & u-\epsilon _{1}\psi
_{2}%
\end{array}%
\right\vert
\nonumber \\
&=&
-\left\vert
\begin{array}{ccc}
\psi _{0}\medskip  & \psi _{1} & \psi _{2} \\
h_{0}\psi _{0} & h_{1}\psi _{1} & h_{2}\psi _{2} \\
\epsilon _{0}\psi _{0} & \epsilon _{1}\psi _{1} & \epsilon _{2}\psi _{2}%
\end{array}%
\right\vert ~.
\end{eqnarray}%
Hence, taking (\ref{G4}) into account, we can show that
\begin{equation} \label{G6}
\psi^C\left[ 2\right]_3=\frac{\left[ \epsilon _{0}\left\{
h_{2}-h_{1}\right\} -\epsilon _{1}\left\{ h_{2}-h_{0}\right\} +\epsilon
_{2}\left\{ h_{1}-h_{0}\right\} \right] \psi _{0}\psi _{1}\psi _{2}}{\left(
h_{1}-h_{0}\right) \psi _{1}\psi _{0}} ~,
\end{equation}%
which obviously implies that $\psi^D[2]_3=\psi^C[2]_3$.
This, as it stands,  is a general proof for a sub-case of our general theorem. On purpose above we have used
different steps than in the proof of our general theorem. The idea behind it is to
demonstrate that in an explicit example we would be only repeating he very same steps as above.
It is therefore sufficient to calculate every time only the right hand side of (\ref{G6}).
The equality $\psi^D[2]_3=\psi^C[2]_3$ is guaranteed by (\ref{G3}) and {\ref{G6}).
We can now apply the results for $\psi^D[2]_3$ by choosing the Ginocchio potential
\begin{equation} \label{G7}
V\left( x\right) =\left\{ -\beta ^{2}\upsilon \left( \upsilon +1\right) +%
\frac{1}{4}\left( 1-\beta ^{2}\right) \left[ 5\left( 1-\beta ^{2}\right)
y^{4}-\left( 7-\beta ^{2}\right) y^{2}+2\right] \right\} \left(
1-y^{2}\right)~,
\end{equation}%
where $y\left( x\right) $ satisfies the following differential equation
\begin{equation} \label{G8}
\frac{dy}{dx}=\left( 1-y^{2}\right) \left[ 1-\left( 1-\beta ^{2}\right) y^{2}%
\right]
\end{equation}
and $\beta$, $\upsilon$ are parameters.

The wave functions of this problem are known to be expressible through Gegenbauer polynomials
$C_n^{(a)}(x)$, namely
\begin{equation} \label{G9}
\psi _{n}=\left( 1-\beta ^{2}\right) ^{\mu _{n}/2}\left[ g\left( y\right) %
\right] ^{-\left( 2\mu _{n}+1\right) /4}C_{n}^{\left( \mu _{n}+1/2\right)
}\left( f\left( y\right) \right)~,
\end{equation}%
where
\begin{equation} \label{G10}
g\left( y\right) =1-\left( 1-\beta ^{2}\right) y^{2}, \,\,\,\,
f\left( y\right) =\frac{\beta y}{\sqrt{g\left( y\right) }}~.
\end{equation}%
The value of $\mu_n$ is connected to the eigenvalue $\epsilon_n$ by $\epsilon_n=-\mu^2_n\beta^4$
such that
\begin{equation} \label{G11}
\mu_n\beta^2 =\sqrt{\beta^2(\upsilon +1/2)^2 +(1-\beta^2)(n+1/2)^2} -(n+1/2)~.
\end{equation}
The first four Gegenbauer polynomials are given as follows
\begin{eqnarray} \label{G12}
C_{0}^{\left( \mu _{0}+1/2\right) }\left( f\left( y\right) \right) &=&1
\nonumber \\
C_{1}^{\left( \mu _{1}+1/2\right) }\left( f\left( y\right) \right) &=&2\left(
\mu _{1}+1/2\right) \left[ f\left( y\right) \right]
\nonumber \\
C_{2}^{\left( \mu _{2}+1/2\right) }\left( f\left( y\right) \right)& = & 2\left(
\mu _{2}+1/2\right) \left( \mu _{2}+3/2\right) \left[ f\left( y\right) %
\right] ^{2}-\left( \mu _{2}+1/2\right)
\nonumber \\
C_{3}^{\left( \mu _{3}+1/2\right) }\left( f\left( y\right) \right)  &=&\frac{%
4}{3}\left( \mu _{3}+1/2\right) \left( \mu _{3}+3/2\right) \left( \mu
_{3}+5/2\right) \left[ f\left( y\right) \right] ^{3}\bigskip  \\
&&-2\left( \mu _{3}+1/2\right) \left( \mu _{3}+3/2\right) \left[ f\left(
y\right) \right]~.
\end{eqnarray}
These functions can be used, in the next step, to compute explicitly the ratios $h_i=\psi^{\prime}_i/\psi_i$.
We obtain
\begin{eqnarray} \label{G13}
h_0 &=& \frac{\psi _{0}^{\prime }}{\psi _{0}}=\frac{\left[ g\left( y\right) \right]
^{\prime }}{g\left( y\right) }=-2\left( 1-\beta ^{2}\right) y\left(
1-y^{2}\right)
\nonumber \\
h_1&=&\frac{\psi _{1}^{\prime }}{\psi _{1}}=\frac{\left[ g\left( y\right) \right]
^{\prime }}{\left[ g\left( y\right) \right] }+\frac{\left[ f\left( y\right) %
\right] ^{\prime }}{\left[ f\left( y\right) \right] }=\frac{\left(
1-y^{2}\right) }{y}\left\{ 1-2\left( 1-\beta ^{2}\right) y^{2}\right\}
\nonumber \\
h_2 &=& \frac{\psi _{2}^{\prime }}{\psi_2}=\left( \frac{\left[ g\left( y\right) \right] ^{\prime }}{%
\left[ g\left( y\right) \right] }+\frac{2\left[ f\left( y\right) \right] %
\left[ f\left( y\right) \right] ^{\prime }}{\left( \left[ f\left( y\right) %
\right] ^{2}-\frac{1}{\left( 2\mu _{2}+3\right) }\right) }\right)~,
\end{eqnarray}
where we have used
\begin{equation} \label{G14}
\frac{\left[ f\left( y\right) %
\right] ^{\prime }}{\left[ f\left( y\right)\right]}=\frac{1}{y}\left(1-y^2 \right)~.
\end{equation}
Noting that the $h_i$ are proportional $(1-y^2)$ and that $h_1-h_0=(1-y^2)/y$
we can insert our results into equation (\ref{G6}) to obtain
\begin{eqnarray} \label{G15}
\psi^C[2]_3 &=&
\left( 1-\beta ^{2}\right) ^{\mu _{2}/2}\left( \mu
_{2}+1/2\right) \left( 2\mu _{2}+3\right) \left[ g\left( y\right) \right]
^{-\left( 2\mu _{2}+1\right) /4}\nonumber\\
& \times &
\left\{(\epsilon_2-\epsilon_0)
\left( \left[ f\left( y\right) \right] ^{2}-%
\frac{1}{\left( 2\mu _{2}+3\right)}\right)
-(\epsilon_1-\epsilon_0)2[f(y)]^2\right\}~.
\end{eqnarray}%
Turning our attention to the potential the superpartner of $V$
in equation (\ref{G7}) it is not difficult to see that
the superpartner is given by
\begin{eqnarray} \label{16}
V^D\left[ 1\right] &=&V^C\left[ 1%
\right] =V-2%
\frac{d^{2}}{dx^{2}}\ln W_{1}
=V-2%
\frac{d}{dr}\frac{\left[ g\left( y\right) \right] ^{\prime }}{g\left(
y\right) }
\nonumber \\
&=&V+4\left( 1-\beta ^{2}\right)\left( 1-3y^{2}\right) ^{2}
\left( 1-y^{2}\right) \left[ 1-\left( 1-\beta ^{2}\right) y^{2}%
\right]~.
\end{eqnarray}
The second Crum iteration yields
\begin{eqnarray} \label{G17}
V^C\left[ 2\right] &=&V-2\frac{d^{2}}{dx^{2}}\ln W_{2}
\nonumber \\
= V&-&2\biggl[\left\{ -2+\left( 1-\beta
^{2}\right) \left[ 5y^{2}-3\right] \right\} +
10y^{2}\left( 1-\beta ^{2}\right) \biggr]
\left( 1-y^{2}\right) \left[ 1-\left( 1-\beta ^{2}\right) y^{2}%
\right]~.
\end{eqnarray}%
To proof that this is equivalent to the second Darboux transformations it is
convenient, as it was the case with the wave functions, to provide first a general proof
for this sub-case. Starting with the definition, it is straightforward to show that
\begin{equation} \label{G18}
\psi^D\left[ 1\right]_2=\left( h_{1}-h_{0}\right) \psi
_{1}~,
\end{equation}%
which leads to
\begin{eqnarray} \label{G19}
V^D\left[ 2\right] &=& V^D\left[ 1\right] -2%
\frac{d}{dr}\frac{(\psi^D\left[ 1\right]_2)^{\prime}}{\psi^D\left[ 1\right]_2}
=V-2\frac{d^{2}}{dr^{2}}\ln \psi_{0}-2%
\frac{d^{2}}{dr^{2}}\ln \left( h_{1}-h_{0}\right) \psi_{1}
\nonumber \\
&=& V-2\frac{d^{2}}{dr^{2}}\left\{ \ln
\left( h_{1}-h_{0}\right) \psi_{0}\psi_{1}\right\}
=V-2\frac{d^{2}}{dr^{2}}\ln W_{2}=V^C\left[ 2\right]~.
\end{eqnarray}
In taking explicit examples we would be only repeating the very same
steps as above.
This demonstration concludes our two examples.

%.........................................................................5555555555555555555555555555

\section{The connection between shape invariance and Crum transformations}

In view of the results of the previous section we can now drop the
distinction between higher order Darboux ($D$) and Crum ($C$)
transformations.

Let $a$ denote a set of parameters in the original potential, i.e.,
\begin{equation}
u=u(x;a)~.
\end{equation}

The condition for shape invariance of $u$ is given by
\begin{equation}
u\left[ 1\right] \left( x;a\right) =u(x;f\left( a\right) )+R\left( f\left(
a\right) \right) ~,
\end{equation}%
where $u \left[1\right](x;a)$ is the first Darboux transform
of the original potential, $f$ transforms $a$ into another set $f\left( a\right) $ and $R\left(
f\left( a\right) \right) $ is a function of the parameters. In the following,
we use the usual notation $a_{m}\equiv f^{m}\left( a\right) $, where $m$
indicates the function $f$ applied $m$ times.

In the preceding section we established an equivalence between higher order
Darboux transformation and the Crum result. Since the shape invariance is
given in terms of the first order Darboux transformation, it is legitimate
to ask if higher order Darboux transformations (Crum transformations) play a
role in the Schr\"{o}dinger equation with shape invariant potentials. As a
first step we will prove the following theorem

\begin{lem}
Under the condition of shape invariance one has
\begin{equation} \label{shapewave}
\psi _{s}\left( x;a_{1}\right) =\psi \left[ 1\right] _{s+1}\left( x;a\right)~.
\end{equation}%
up to a multiplicative constant and
\begin{equation}
\lambda _{s}\left( a_{1}\right) +R\left( a_{1}\right) =\lambda _{s+1}\left(
a\right)~.   \label{shapevalue}
\end{equation}
In the above $\psi _{s}\left( x;a \right)$ denotes the eigenfunction to the Hamiltonian
with the potential $u (x;a)$ with the eigenvalue $\lambda_s$.
\end{lem}

These results are not new. But since we will make use of them, we offer here
a short proof. We start with initial Sturm-Liouville problem
\begin{equation}
\left( -\frac{d^{2}}{dx^{2}}+u\left( x;a\right) \right) \psi _{s}\left(
x;a\right) =\lambda _{s}\left( a\right) \psi _{s}\left( x;a\right)
\label{coor}
\end{equation}%
and
\begin{equation}
\left( -\frac{d^{2}}{dx^{2}}+u\left[ 1\right] \left( x;a\right) \right) \psi %
\left[ 1\right] _{s}\left( x;a\right) =\lambda _{s}\left( a\right) \psi %
\left[ 1\right] _{s}\left( x;a\right) \quad s>1 ~.  \label{partn}
\end{equation}%
Equation (\ref{coor}) is valid for any $a$, hence we may write
\begin{equation}
\left( -\frac{d^{2}}{dx^{2}}+u\left( x;a_{1}\right) \right) \psi _{s}\left(
x;a_{1}\right) =\lambda _{s}\left( a_{1}\right) \psi _{s}\left(
x;a_{1}\right)
\end{equation}%
and add $R\left( f\left( a\right) \right) \psi _{s}\left( x;f\left( a\right)
\right) $ on both sides implying the following identity
\begin{equation}
\left( -\frac{d^{2}}{dx^{2}}+u\left( x;a_{1}\right) +R\left( a_{1}\right)
\right) \psi _{s}\left( x;a_{1}\right) =\left( \lambda _{s}\left(
a_{1}\right) +R\left( a_{1}\right) \right) \psi _{s}\left( x;a_{1}\right)~.
\end{equation}%
Due to the shape invariance condition (for the sake of formulating the next Lemma
we can say that $u\left[ 1\right] $ and $u$ are pairwise shape invariant)
this becomes
\begin{equation}
\left( -\frac{d^{2}}{dx^{2}}+u\left[ 1\right] \left( x;a\right) \right) \psi
_{s}\left( x;a_{1}\right) =\left( \lambda _{s}\left( a_{1}\right) +R\left(
a_{1}\right) \right) \psi _{s}\left( x;a_{1}\right)~.
\end{equation}%
Without loss of generality, the spectrum can be ordered as $\lambda
_{1}<\lambda _{2}<\lambda _{3}<...$. Hence, $\{\lambda _{s}(a)\}$, $\{\lambda
_{s}(a_{1})\}$ and $\{\lambda _{s}(a_{1})+R(a_{1})\}$ are similarly ordered
sets. $\psi \left[ 1\right] _{s+1}$ is then an eigenfunction to the ordered
spectrum $\lambda _{2}<\lambda _{3}<...$. We can conclude that up to a
multiplicative factor
\begin{equation}
\psi _{s}\left( x;a_{1}\right) =\psi \left[ 1\right] _{s+1}(x,\,a)
\label{shapefunc}
\end{equation}%
and
\begin{equation}
\lambda _{s}\left( a_{1}\right) +R\left( a_{1}\right) =\lambda _{s+1}\left(
a\right)~.   \label{shapevalue2}
\end{equation}

In preparation of the main theorem of this section we prove the next Lemma.

\begin{lem}
By virtue of the the above Lemma and under the condition that $u$ and $u%
\left[ 1\right] $ are pairwise shape invariant, $u\left[ 1\right] $, $u\left[
2\right] $ are also pairwise shape invariant i.e.
\begin{equation}
u\left[ 2\right] \left( x;a\right) =u\left[ 1\right] \left( x;a_{1}\right)
+R\left( a_{1}\right)~.
\end{equation}
\end{lem}

The proof can be done in two steps.

\begin{enumerate}
\item The condition of shape invariance and the definition of the Darboux
transformation allows us to write
\begin{equation}
u\left( x;a\right) -2\frac{d}{dx}\frac{\psi _{1}^{\prime }\left( x;a\right)
}{\psi _{1}\left( x;a\right) }=u\left[ 1\right] \left( x;a\right)
=u(x;a_{1})+R\left( a_{1}\right)
\end{equation}%
which remains valid if we replace $a$ by $a_{1}$, i.e.,
\begin{equation}
u\left( x;a_{1}\right) -2\frac{d}{dx}\frac{\psi _{1}^{\prime }\left(
x;a_{1}\right) }{\psi _{1}\left( x;a_{1}\right) }=u\left[ 1\right] \left(
x;a_{1}\right) =u(x;a_{2})+R\left( a_{2}\right)~.
\end{equation}%
Hence, we easily obtain
\begin{equation}
u\left( x;a_{1}\right) =u\left[ 1\right] \left( x;a_{1}\right) +2\frac{d}{dx}%
\frac{\psi _{1}^{\prime }\left( x;a_{1}\right) }{\psi _{1}\left(
x;a_{1}\right) }~.  \label{ufa}
\end{equation}

\item Applying the Darboux transformation (\ref{132}), once again on $u\left[
1\right] \left( x;a\right) $,  gives
\begin{equation}
u\left[ 2\right] \left( x;a\right) =u\left[ 1\right] \left( x;a\right) -2%
\frac{d}{dx}\frac{\psi ^{\prime }\left[ 1\right] _{2}\left( x;a\right) }{%
\psi \left[ 1\right] _{2}\left( x;a\right) }~.
\end{equation}%
On the other hand,
using the shape invariance condition leads to
\begin{equation}
u\left[ 2\right] \left( x;a\right) =u\left( x;a_{1}\right) +R\left(
a_{1}\right) -2\frac{d}{dx}\frac{\psi ^{\prime }\left[ 1\right] _{2}\left(
x;a\right) }{\psi \left[ 1\right] _{2}\left( x;a\right) }~.
\end{equation}%
The result in the first step of the proof, (\ref{ufa}), can be used to
derive the following equation:
\begin{equation}
u\left[ 2\right] \left( x;a\right) =u\left[ 1\right] \left( x;a_{1}\right)
+R\left( a_{1}\right) +2\frac{d}{dx}\left\{ \frac{\psi _{1}^{\prime }\left(
x;a_{1}\right) }{\psi _{1}\left( x;a_{1}\right) }-\frac{\psi ^{\prime }\left[
1\right] _{2}\left( x;a\right) }{\psi \left[ 1\right] _{2}\left( x;a\right) }%
\right\}~.
\end{equation}%
If we now apply (\ref{shapefunc}) from Lemma V.1 for $s=1$, i.e.,
\begin{equation}
\psi \left[ 1\right] _{2}\left( x;a\right) =\psi _{1}\left( x;a_{1}\right)
\label{wavef}
\end{equation}%
we obtain the desired final expression which we wanted to prove, namely,
\begin{equation}
u\left[ 2\right] \left( x;a\right) =u\left[ 1\right] \left( x;a_{1}\right)
+R\left( a_{1}\right)~.   \label{shapepot}
\end{equation}
\end{enumerate}

For the sake of a more compact notation of the
properties of the potential and wave functions,
let us now call the property (\ref{shapefunc}) \textit{shape invariance for
eigenfunctions} (or better \textit{the two eigenfunctions involved are
pairwise shape invariant}) and (\ref{shapevalue}) \textit{shape invariance
for the eigenvalues}. Note that the shape invariance of the wave functions follows from the
shape invariance of the potentials. From the shape invariance of the
eigenfunction we can, in turn, conclude that the next two pairs of Darboux
transformations of the potential are pairwise shape invariant. One is led to
the conjecture that the chain continues: from Lemma V.2 one can show
that the next pair of higher order Darboux transformations of eigenfunctions
are also pairwise shape invariant, from which it follows that the next
higher order pair of Darboux transformed potentials is also pairwise shape
invariant. Indeed, we can prove the following theorem extending hereby the notion of shape invariance.

\begin{thm}
All neigbouring higher order Darboux transformed potentials and
eigenfunctions are pairwise shape invariant. This is to say,
\begin{equation}
u\left[ k\right] \left( x;a\right) =u\left[ k-1\right] \left(
x;a_{1})+R\left( a_{1}\right) \right)   \label{one1}
\end{equation}%
and
\begin{equation}
\psi \left[ k\right] _{s+1}\left( x;a\right) =\psi \left[ k-1\right]
_{s}\left( x;a_{1}\right)~,   \label{two2}
\end{equation}%
up to a multiplicative factor. In more detail, (\ref{one1}) implies (\ref%
{two2}) which, in turn, implies
\begin{equation}
u\left[ k+1\right] \left( x;a\right) =u\left[ k\right] \left( x;a_{1}\right)
+R\left( a_{1}\right)~.   \label{hipoinducc2}
\end{equation}
\end{thm}

The proof proceeds via induction whose first step consists in Lemma V.2 and
Lemma V.1 or in (\ref{shapefunc}, \ref{shapepot}). We assume the hypothesis
of the induction to be [(\ref{one1}) $\Rightarrow $ (\ref{two2})]. This is
sufficient since we start with the original shape invariance condition for
potentials and the first step of induction is presented in Lemma V.1 and
Lemma V.2. We have to show that under this condition
\begin{equation}
\psi \left[ k+1\right] _{s+1}\left( x;a\right) =\psi \left[ k\right]
_{s}\left( x;a_{1}\right)
\end{equation}%
holds, from which, in turn,
\begin{equation}
u\left[ k+2\right] \left( x;a\right) =u\left[ k+1\right] \left(
x;a_{1}\right) +R\left( a_{1}\right)
\end{equation}%
follows.

\begin{enumerate}
\item We have
\begin{equation}
\left( -\frac{d^2}{dx^2} + u\left[ k\right] \left( x;a\right) \right) \psi \left[ k\right]
_{s}\left( x;a\right) =\lambda _{s}\left( a\right) \psi \left[ k\right]
_{s}\left( x;a\right) \quad s>k ~.
\end{equation}

\item Since the above equation is valid for any $a$, it is also valid when
$a$ is replaced by $f(a)$. If we add $R\left( a_{1}\right) \psi \left[ k\right]
_{s}\left( x;a_{1}\right) $ on both sides and make use of the induction
hypothesis we arrive, for $s>k$, at
\begin{equation}
\left( -\frac{d^{2}}{dx^{2}}+u\left[ k+1\right] \left( x;a\right) \right)
\psi \left[ k\right] _{s}\left( x;a_{1}\right) =\left( \lambda _{s}\left(
a_{1}\right) +R\left( a_{1}\right) \right) \psi \left[ k\right] _{s}\left(
x;a_{1}\right)~.
\end{equation}%
Equation (\ref{shapevalue}) then tells us that,
\begin{equation}
\psi \left[ k\right] _{s}\left( x;a_{1}\right) =\psi \left[ k+1\right]
_{s+1}\left( x;a\right)~,    \label{shapefun1}
\end{equation}%
up to a multiplicative factor.

\item By definition we have
\begin{equation}
u\left[ k+1\right] \left( x;a\right) =u\left[ k\right] \left( x;a\right) -2%
\frac{d}{dx}\left( \frac{\psi ^{\prime }\left[ k\right] _{k+1}\left(
x;a\right) }{\psi \left[ k\right] _{k+1}\left( x;a\right) }\right)
\end{equation}%
for any $a$. Hence also:
\begin{equation}
u\left[ k+1\right] \left( x;a_{1}\right) =u\left[ k\right] \left(
x;a_{1}\right) -2\frac{d}{dx}\left( \frac{\psi ^{\prime }\left[ k\right]
_{k+1}\left( x;a_{1}\right) }{\psi \left[ k\right] _{k+1}\left(
x;a_{1}\right) }\right)~.
\end{equation}%
Taking $u\left[ k\right] \left( x;a_{1}\right) $ from this equation and
inserting the result in the induction hypothesis, one can easily show that
\begin{equation}
u\left[ k+1\right] \left( x;a\right) =u\left[ k+1\right] \left(
x;a_{1}\right) +2\frac{d}{dx}\left( \frac{\psi ^{\prime }\left[ k\right]
_{k+1}\left( x;a_{1}\right) }{\psi \left[ k\right] _{k+1}\left(
x;a_{1}\right) }\right) +R\left( a_{1}\right)~.
\end{equation}

\item Again per definition we know that
\begin{equation}
u\left[ k+2\right] \left( x;a\right) =u\left[ k+1\right] \left( x;a\right) -2%
\frac{d}{dx}\left( \frac{\psi ^{\prime }\left[ k+1\right] _{k+2}\left(
x;a\right) }{\psi \left[ k+1\right] _{k+2}\left( x;a\right) }\right)~.
\end{equation}

\item Combining the last two equations yields
\begin{equation}
u\left[ k+2\right] \left( x;a\right) =u\left[ k+1\right] \left(
x;a_{1}\right) +R\left( a_{1}\right) +2\frac{d}{dx}\left( \frac{\psi
^{\prime }\left[ k\right] _{k+1}\left( x;a_{1}\right) }{\psi \left[ k\right]
_{k+1}\left( x;a_{1}\right) }-\frac{\psi ^{\prime }\left[ k+1\right]
_{k+2}\left( x;a\right) }{\psi \left[ k+1\right] _{k+2}\left( x;a\right) }%
\right)~.
\end{equation}

\item The last step consists in using the already established result (\ref{shapefun1}%
) to obtain
\begin{equation}
u\left[ k+2\right] \left( x;a\right) =u\left[ k+1\right] \left(
x;a_{1}\right) +R\left( a_{1}\right)~,
\end{equation}
\end{enumerate}

which completes the proof.

The shape invariance condition (more accurately, the shape invariance
between $u$ and $u[1]$) allows one to define a new Hamiltonian of the order $%
s$:
\begin{equation}
H^{SI}_s\equiv -\frac{d^{2}}{dx^{2}}+u(x;a_{s})+\sum_{k=1}^{s}R\left(
a_{k}\right)~.   \label{SI}
\end{equation}%
Note that this definition makes no reference to higher order Darboux (or Crum)
transformations. However, by virtue of the Theorem V.3 we can iterate
\begin{eqnarray}
H^{SI}_s &=&-\frac{d^{2}}{dx^{2}}+u[1](x;a_{s-1})+\sum_{k=1}^{s-1}R\left(
a_{k}\right)   \notag \\
&=&-\frac{d^{2}}{dx^{2}}+u[2](x;a_{s-2})+\sum_{k=1}^{s-2}R\left(
a_{k}\right)   \notag \\
&=&...=-\frac{d^{2}}{dx^{2}}+u[s](x;a)=H^{D}_s ~.
\end{eqnarray}%
Hence, in view of the above and the Theorem III.1 we can state as a corollary

\begin{cor}
Under the condition of shape invariance all three transformations are equal,
i.e.,
\begin{equation}
H^{SI}_s=H^{D}_s=H^{C}_s ~.  \label{allfine}
\end{equation}
\end{cor}

\section{example}

We will continue here with our example of the Morse potential, now however
emphasizing the aspect of shape invariance around Lemmas V.1-V.2 and Theorem
V.3. Indeed, the Morse potential is shape invariant. One defines the action of $f$
as
\begin{equation}
f\left( A\right)\equiv A_{1} =A-\frac{\alpha }{\sqrt{2}}
\end{equation}%
in accordance with the notation in (\ref{extray}). $R$ is identified with
\begin{equation}
R\left( A_{1}\right) =2\left( A^{2}-A_{1}^{2}\right)~.
\end{equation}
Note that
\begin{eqnarray}
\psi _{1}\left( x;A_{1}\right)  &=&c_{1}\left( A_{1}\right) \left[ {\rm
sech}\left( \alpha x\right) \right] ^{\frac{\sqrt{2}A_1 }{%
\alpha }}=c_{1}\left( A_{1}\right) \left[ {\rm sech}\left( \alpha
x\right) \right] ^{\frac{\sqrt{2}A}{\alpha }-1}\medskip   \notag \\
&=&c\cosh \left( \alpha x\right) \psi _{1}\left( x;A\right)
\end{eqnarray}%
immediately leads to
\begin{equation}
\psi \left[ 1\right] _{2}\left( x;A\right) =\psi _{1}\left(x;A_1 \right)
\label{psi12shape}
\end{equation}%
which is valid up to a multiplicative constant. One also verifies that
the equality below
\begin{equation}
\psi _{2}\left( x;A_{1}\right) =c_{2}\left( A_{1}\right) \sinh \left( \alpha
x\right) \psi _{1}\left( x;A_{1}\right)
\end{equation}%
\begin{equation}
=c_{2}\left( A_{1}\right) \sinh \left( \alpha x\right) \cosh \left( \alpha
x\right) \psi _{1}\left( x;A\right)
\end{equation}%
together with (\ref{extra43}) has as a consequence the following
identity (again up to a constant multiplicative value)%
\begin{equation} \label{extrax2}
\psi \left[ 1\right] _{3}(x;A)=\psi _{2}\left( x;A_{1}\right)~.
\end{equation}%
Equations (\ref{psi12shape}) and (\ref{extrax2}) are explicit examples of the result
(\ref{shapewave}) in Lemma  V.1.
Regarding the eigenvalues, i.e. the property (\ref{shapevalue}) in the same
Lemma, let us first note that another compact notation for equation (\ref{eigenvaluesx})
is
\begin{equation}
\lambda _{n}\left( A_{1}\right) =2\left( A_{1}^{2}-A_{n}^{2}\right)
\end{equation}%
which leaves us with the identity
\begin{equation}
\lambda _{n}\left( A_{1}\right) +R\left( A_{1}\right) =\lambda _{n+1}\left(
A\right)~,
\end{equation}%
as it should be according to Lemma V.1. Finally, we can also give
explicit examples regarding Theorem V.1.
Due to the results from section 5, we can write,
\begin{eqnarray}
u\left[ 1\right] \left( x;A_{1}\right) +R\left( A_{1}\right)  &=&2\left[
A_{1}^{2}-A_{1}\left( A_{1}-\frac{\alpha }{\sqrt{2}}\right) \mathrm{sech}%
{}^{2}\left( \alpha x\right) \right] +2\left( A^{2}-A_{1}^{2}\right)   \notag
\\
&=&2\left[ A^{2}-A_{1}A_{2}\mathrm{sech}{}^{2}\left( \alpha x\right) \right]
=u\left[ 2\right] \left( x;A\right)~.
\end{eqnarray}%
This demonstrates in an explicit example the result (\ref{one1}) from Theorem V.1.
Last but not least, one sees that equation (\ref{extra42}) can be written as
\begin{equation}
\psi \left[ 1\right] _{2}\left( x;A_{1}\right) =c_{2}\left( A_{1}\right)
\alpha \cosh \left( \alpha x\right) \psi _{1}\left( x;A_{1}\right) ~,
\end{equation}%
which, according to (\ref{psi12shape}) can be cast into the following form:
\begin{equation*} \label{last3}
\psi \left[ 1\right] _{2}\left( x;A_{1}\right) =\frac{c_{2}\left(
A_{1}\right) }{c}\alpha \cosh \left( \alpha x\right) \psi \left[ 1\right]
_{2}\left( x;A\right) =\frac{c_{2}\left( A_{1}\right) c_{2}\left( A\right) }{%
c}\alpha ^{2}\cosh ^{2}\left( \alpha x\right) \psi _{1}\left( x;A\right)
\end{equation*}%
\begin{equation}
=\bar{c}\psi \left[ 2\right] _{3}\left( x;A\right)
\end{equation}%
with $\bar{c}$ a constant. To arrive at the last result
we have used (\ref{last2}) from which one can also
determine the constant $\bar{c}$ in terms of $\lambda_3$, $c_2(A)$,
$c_2(A_1)$ and $c_3(A)$.
Obviously, the above equation falls into the
category of explicit examples of (\ref{two2}). Note that in none of the above
examples we have used the actual Lemmas or Theorems to be exemplified (as it should be if an example
carries some meaning).
%.......................777777777777777777777777777777777777777777777777777

\section{Conclusions}

In the present work, we have clarified the relations between the Darboux and
Crum transformations. We have shown that the latter can be reached
iteratively by higher order Darboux transformations. This is valid for the
potential as well as the eigenfunctions. If we subject the potential to the
condition of shape invariance, another transform (not making use of Crum
transformation for $n >1)$ is possible (equation (\ref{SI})). We prove that
this is also equivalent to the Crum transform. The main steps of this proof
involved establishing (\ref{one1}), (\ref{two2}) and (\ref{shapevalue}). The
first result, namely (\ref{one1}), is a generalization of the original shape invariance condition. Note
that (\ref{two2}) and (\ref{shapevalue}) could be called shape invariance
for the wave functions and eigenvalues. The results of the paper help to
understand the relation between the different transformations of the
Hamiltonian operator. Indeed, in view of our results, one could say that the
Crum transformation which appears much more complex than the original
Darboux result is essentially an iterative higher order Darboux transformation and
therefore not more complex than the former.

%.....................................................888888888888888888888888888888

\section{Appendix A: An application of Jacobi Theorem}

The identity $W\left( W_{n},W_{n-1,s}\right) =W_{ns}W_{n-1}$ has been used
by Crum in the proof of his theorem. We have also made use of it several
times in the present paper. It therefore makes sense to sketch a proof of
the same.

Let us first establish some notations and definitions. Let $A=[a_{ij}]$ be a
$n\times n$ matrix. The determinant of $A$ will be denoted by $\left|
A\right|$ as usual. We call the minor $M_{r}$, the determinant obtained by
retaining from $A$ the $r$ lines $i_{1}$, $i_{2},$ ..., $i_{r}$ and the $r$
columns $k_{1}$, $k_{2},$ ..., $k_{r}.$ One defines the complement of the
minor $M_{r}$ as the determinant obtained from $A$ by dropping the $r$ lines
$i_{1}$, $i_{2},$ ..., $i_{r}$ and the $r$ columns $k_{1}$, $k_{2},$ ..., $%
k_{r}$. This complement will be denoted by $M_{r}^{c}$ . One then defines $%
M^{\left( r\right) }$%
\begin{equation}
M^{\left( r\right) }=\left( -1\right)
^{i_{1}+i_{2}+...+i_{r}+k_{1}+k_{2}+...+k_{r}}M_{r}^{c}~.  \label{201}
\end{equation}
Furthermore, let $\Delta $ be the matrix of the cofactors of $A:$%
\begin{equation}
\Delta =\left|
\begin{array}{cccc}
A_{11} & A_{12} & \cdots & A_{1n} \\
A_{21} & A_{22} & \cdots & A_{2n} \\
\vdots & \vdots & \ddots & \vdots \\
A_{n1} & A_{n2} & \cdots & A_{nn}%
\end{array}
\right|
\end{equation}
and $M_{r}$, $M_{r}^{\prime }$ the minors of $\left| A\right| $ and $\Delta $%
, respectively.

\begin{Jacobi}
With these notations, the theorem of Jacobi asserts that
\begin{equation}
M_{r}^{\prime }=\left| A\right| ^{r-1}M^{\left( r\right) }~.  \label{jacobi}
\end{equation}
\end{Jacobi}

Before proceeding we make a small diversion to an example of the application
of the above theorem starting with a Wronskian composed of $\psi _{1},$ $%
\psi _{2},$ $\psi _{3},$ $\psi _{s}$, i.e.,
\begin{equation}
\left| A\right| \equiv W_{3,s}=\left|
\begin{array}{cccc}
\psi _{1} & \psi _{2} & \psi _{3} & \psi _{s} \\
\psi _{1}^{\prime } & \psi _{2}^{\prime } & \psi _{3}^{\prime } & \psi
_{s}^{\prime } \\
\psi _{1}^{\prime \prime } & \psi _{2}^{\prime \prime } & \psi _{3}^{\prime
\prime } & \psi _{s}^{\prime \prime } \\
\psi _{1}^{\prime \prime \prime } & \psi _{2}^{\prime \prime \prime } & \psi
_{3}^{\prime \prime \prime } & \psi _{s}^{\prime \prime \prime }%
\end{array}
\right|~.
\end{equation}
The matrix of the cofactors is then given by $\medskip \medskip $%
\begin{eqnarray}
& &\Delta =  \notag \\
& &\left|
\begin{array}{cccc}
+\left|
\begin{array}{ccc}
\psi _{2}^{\prime } & \psi _{3}^{\prime } & \psi _{s}^{\prime } \\
\psi _{2}^{\prime \prime } & \psi _{3}^{\prime \prime } & \psi _{s}^{\prime
\prime } \\
\psi _{2}^{\prime \prime \prime } & \psi _{3}^{\prime \prime \prime } & \psi
_{s}^{\prime \prime \prime }%
\end{array}
\right| \medskip \medskip \quad & -\left|
\begin{array}{ccc}
\psi _{1}^{\prime } & \psi _{3}^{\prime } & \psi _{s}^{\prime } \\
\psi _{1}^{\prime \prime } & \psi _{3}^{\prime \prime } & \psi _{s}^{\prime
\prime } \\
\psi _{1}^{\prime \prime \prime } & \psi _{3}^{\prime \prime \prime } & \psi
_{s}^{\prime \prime \prime }%
\end{array}
\right| \quad & +\left|
\begin{array}{ccc}
\psi _{1}^{\prime } & \psi _{2}^{\prime } & \psi _{s}^{\prime } \\
\psi _{1}^{\prime \prime } & \psi _{2}^{\prime \prime } & \psi _{s}^{\prime
\prime } \\
\psi _{1}^{\prime \prime \prime } & \psi _{2}^{\prime \prime \prime } & \psi
_{s}^{\prime \prime \prime }%
\end{array}
\right| \quad & -\left|
\begin{array}{ccc}
\psi _{1}^{\prime } & \psi _{2}^{\prime } & \psi _{3}^{\prime } \\
\psi _{1}^{\prime \prime } & \psi _{2}^{\prime \prime } & \psi _{3}^{\prime
\prime } \\
\psi _{1}^{\prime \prime \prime } & \psi _{2}^{\prime \prime \prime } & \psi
_{3}^{\prime \prime \prime }%
\end{array}
\right| \\
-\left|
\begin{array}{ccc}
\psi _{2} & \psi _{3} & \psi _{s} \\
\psi _{2}^{\prime \prime } & \psi _{3}^{\prime \prime } & \psi _{s}^{\prime
\prime } \\
\psi _{2}^{\prime \prime \prime } & \psi _{3}^{\prime \prime \prime } & \psi
_{s}^{\prime \prime \prime }%
\end{array}
\right| \medskip \medskip \quad & +\left|
\begin{array}{ccc}
\psi _{1} & \psi _{3} & \psi _{s} \\
\psi _{1}^{\prime \prime } & \psi _{3}^{\prime \prime } & \psi _{s}^{\prime
\prime } \\
\psi _{1}^{\prime \prime \prime } & \psi _{3}^{\prime \prime \prime } & \psi
_{s}^{\prime \prime \prime }%
\end{array}
\right| \quad & -\left|
\begin{array}{ccc}
\psi _{1} & \psi _{2} & \psi _{s} \\
\psi _{1}^{\prime \prime } & \psi _{2}^{\prime \prime } & \psi _{s}^{\prime
\prime } \\
\psi _{1}^{\prime \prime \prime } & \psi _{2}^{\prime \prime \prime } & \psi
_{s}^{\prime \prime \prime }%
\end{array}
\right| \quad & +\left|
\begin{array}{ccc}
\psi _{1} & \psi _{2} & \psi _{3} \\
\psi _{1}^{\prime \prime } & \psi _{2}^{\prime \prime } & \psi _{3}^{\prime
\prime } \\
\psi _{1}^{\prime \prime \prime } & \psi _{2}^{\prime \prime \prime } & \psi
_{3}^{\prime \prime \prime }%
\end{array}
\right| \\
+\left|
\begin{array}{ccc}
\psi _{2} & \psi _{3} & \psi _{s} \\
\psi _{2}^{\prime } & \psi _{3}^{\prime } & \psi _{s}^{\prime } \\
\psi _{2}^{\prime \prime \prime } & \psi _{3}^{\prime \prime \prime } & \psi
_{s}^{\prime \prime \prime }%
\end{array}
\right| \medskip \medskip \quad & -\left|
\begin{array}{ccc}
\psi _{1} & \psi _{3} & \psi _{s} \\
\psi _{1}^{\prime } & \psi _{3}^{\prime } & \psi _{s}^{\prime } \\
\psi _{1}^{\prime \prime \prime } & \psi _{3}^{\prime \prime \prime } & \psi
_{s}^{\prime \prime \prime }%
\end{array}
\right| \quad & +\left|
\begin{array}{ccc}
\psi _{1} & \psi _{2} & \psi _{s} \\
\psi _{1}^{\prime } & \psi _{2}^{\prime } & \psi _{s}^{\prime } \\
\psi _{1}^{\prime \prime \prime } & \psi _{2}^{\prime \prime \prime } & \psi
_{s}^{\prime \prime \prime }%
\end{array}
\right| \quad & -\left|
\begin{array}{ccc}
\psi _{1} & \psi _{2} & \psi _{3} \\
\psi _{1}^{\prime } & \psi _{2}^{\prime } & \psi _{3}^{\prime } \\
\psi _{1}^{\prime \prime \prime } & \psi _{2}^{\prime \prime \prime } & \psi
_{3}^{\prime \prime \prime }%
\end{array}
\nonumber \right| \\
-\left|
\begin{array}{ccc}
\psi _{2} & \psi _{3} & \psi _{s} \\
\psi _{2}^{\prime } & \psi _{3}^{\prime } & \psi _{s}^{\prime } \\
\psi _{2}^{\prime \prime } & \psi _{3}^{\prime \prime } & \psi _{s}^{\prime
\prime }%
\end{array}
\right| \quad & +\left|
\begin{array}{ccc}
\psi _{1} & \psi _{3} & \psi _{s} \\
\psi _{1}^{\prime } & \psi _{3}^{\prime } & \psi _{s}^{\prime } \\
\psi _{1}^{\prime \prime } & \psi _{3}^{\prime \prime } & \psi _{s}^{\prime
\prime }%
\end{array}
\right| \quad & -\left|
\begin{array}{ccc}
\psi _{1} & \psi _{2} & \psi _{s} \\
\psi _{1}^{\prime } & \psi _{2}^{\prime } & \psi _{s}^{\prime } \\
\psi _{1}^{\prime \prime } & \psi _{2}^{\prime \prime } & \psi _{s}^{\prime
\prime }%
\end{array}
\right| \quad & +\left|
\begin{array}{ccc}
\psi _{1} & \psi _{2} & \psi _{3} \\
\psi _{1}^{\prime } & \psi _{2}^{\prime } & \psi _{3}^{\prime } \\
\psi _{1}^{\prime \prime } & \psi _{2}^{\prime \prime } & \psi _{3}^{\prime
\prime }%
\end{array}
\right|%
\end{array}
\right|.
\end{eqnarray}
We choose as lines and columns: $\left( i_{1},i_{2}\right) =\left(
3,4\right) =\left( k_{1},k_{2}\right) $ . Applying the Jacobi theorem gives
us$\bigskip $%
\begin{equation}
\left|
\begin{array}{cc}
+\left|
\begin{array}{ccc}
\psi _{1} & \psi _{2} & \psi _{s} \\
\psi _{1}^{\prime } & \psi _{2}^{\prime } & \psi _{s}^{\prime } \\
\psi _{1}^{\prime \prime \prime } & \psi _{2}^{\prime \prime \prime } & \psi
_{s}^{\prime \prime \prime }%
\end{array}
\right| \bigskip \quad & -\left|
\begin{array}{ccc}
\psi _{1} & \psi _{2} & \psi _{3} \\
\psi _{1}^{\prime } & \psi _{2}^{\prime } & \psi _{3}^{\prime } \\
\psi _{1}^{\prime \prime \prime } & \psi _{2}^{\prime \prime \prime } & \psi
_{3}^{\prime \prime \prime }%
\end{array}
\right| \\
-\left|
\begin{array}{ccc}
\psi _{1} & \psi _{2} & \psi _{s} \\
\psi _{1}^{\prime } & \psi _{2}^{\prime } & \psi _{s}^{\prime } \\
\psi _{1}^{\prime \prime } & \psi _{2}^{\prime \prime } & \psi _{s}^{\prime
\prime }%
\end{array}
\right| \quad & +\left|
\begin{array}{ccc}
\psi _{1} & \psi _{2} & \psi _{3} \\
\psi _{1}^{\prime } & \psi _{2}^{\prime } & \psi _{3}^{\prime } \\
\psi _{1}^{\prime \prime } & \psi _{2}^{\prime \prime } & \psi _{3}^{\prime
\prime }%
\end{array}
\right|%
\end{array}
\right| =W_{3,s}\left|
\begin{array}{cc}
\psi _{1} & \psi _{2} \\
\psi _{1}^{\prime } & \psi _{2}^{\prime }%
\end{array}
\right|.
\end{equation}
Using Lemma II.1 the left hand side takes the form
\begin{equation}
\left|
\begin{array}{cc}
\frac{d}{dx}W_{2,s}\bigskip & \frac{d}{dx}W_{3} \\
W_{2,s} & W_{3}%
\end{array}
\right|
\end{equation}
such that we can write
\begin{equation}
\left|
\begin{array}{cc}
\frac{d}{dx}W_{2,s}\bigskip & -\frac{d}{dx}W_{3} \\
-W_{2,s} & W_{3}%
\end{array}
\right| =W_{3,s}W_{2} ~.
\end{equation}
Explicitly, this implies the following equality
\begin{equation}
W_{3,s}W_{2}=W_{3}\frac{d}{dx}W_{2,s}-W_{2,s}\frac{d}{dx}W_{3}=W\left(
W_{3},W_{2,s}\right)~.
\end{equation}
The proof of the general case does not require any new procedure and follows
essentially the steps outlined in the example. Let $W_{ns}$ be the Wronskian
of the $n+1$ functions $\psi _{1},\ldots ,\psi _{n},\psi _{s}$,% ,
namely,
\begin{equation}
\left| A\right| =W_{n,s}=\left|
\begin{array}{ccccc}
\psi _{1} & \psi _{2} & ... & \psi _{n} & \psi _{s} \\
\psi _{1}^{\prime } & \psi _{2}^{\prime } & ... & \psi _{n}^{\prime } & \psi
_{s}^{\prime } \\
\vdots & \vdots & \vdots & \vdots & \vdots \\
\psi _{1}^{\left( n-1\right) } & \psi _{2}^{\left( n-1\right) } & ... & \psi
_{n}^{\left( n-1\right) } & \psi _{s}^{\left( n-1\right) } \\
\psi _{1}^{\left( n\right) } & \psi _{2}^{\left( n\right) } & ... & \psi
_{n}^{\left( n\right) } & \psi _{s}^{\left( n\right) }%
\end{array}
\right|
\end{equation}
and $\Delta $ the matrix of the cofactors of $W_{n,s}$ . We would like to
apply the Jacobi theorem for the choice
\begin{equation}
\left( i_{1},i_{2}\right) =\left( n,n+1\right) =\left( k_{1},k_{2}\right)
\end{equation}
such that $r=2$. In this case we need
\begin{equation}
M_{r}^{\prime }=\left|
\begin{array}{ccc}
+W_{n-1,s}^{\prime } &  & -W_{n}^{\prime } \\
&  &  \\
-W_{n-1,s} &  & +W_{n}%
\end{array}
\right| =W\left( W_{n},W_{n-1,s}\right)~,
\end{equation}
where we have used explicitly the result of Lemma II.1. Clearly, we have,
\begin{equation}
M^{\left( r\right) }=W_{n-1}~,
\end{equation}
such that the Jacobi theorem for the Wronskian $A$ can be stated as
\begin{equation}
W\left( W_{n},W_{n-1,s}\right) =W_{ns}W_{n-1}~,
\end{equation}
which proves Lemma 1. %\begin{thebibliography}{salle}
%%%%%%%%%%%%%%%%%%%%%%%%%%%%%%%%%%%%%%%%%%%%%%%%%%%%%%%%%%%%%%%%%%%%%%%%%%%%%%%%%%%%%%%%%%%%%%%%%%%%%%%%%%%%%%%%%%%%%%%%%%%%
%\usepackage{seart}


\begin{thebibliography}{99}
\bibitem{Nicolai} H. Nicolai, J. Phys. \textbf{A9} 1497, (1976)

\bibitem{Witten} E. Witten, Nucl. Phys. \textbf{B188} 513, (1981)

\bibitem{Infeld} L. Infeld and T. E. Hull, Rev. Mod. Phys. \textbf{23}
21,
(1951)

\bibitem{Darboux} G. Darboux, C. R. Acad. Sc. Paris, \textbf{94} 1456, (1882)

\bibitem{Crum} M. Crum, Quart. J. Math. Oxford \textbf{6} 121, (1955)

\bibitem{GM} M. Gel'fand and B. M. Levitan, Am. Math. Soc. Transl \textbf{1} 253,
(1951)

\bibitem{AM} B. Abraham and H. E. Moses, Phys. Rev. \textbf{A22} 1333, (1980)

\bibitem{Deift1} P. A. Deift, Duke Math J. \textbf{45} 267, (1978)

\bibitem{Pursey1} M. Luban and D. L. Pursey, Phys. Rev. \textbf{D33} 431, (1986)

\bibitem{M} V. A. Marchenko, Dkl. Akad. Nauk. SSSR, \textbf{104} 695, (1955)
\bibitem{Pursey2} D. L. Pursey, Phys. Rev. \textbf{D33} 1048, (1986)

\bibitem{Gendenshtein} L. Gendenshtein, Pis'ma Zh. Eksp. Teor. Fiz. \textbf{%
38} 299, (1983) [JETP Lett. \textbf{38} 356, (1983)]

\bibitem{Barklay} D.T. Barklay, R. Dutt, A. Gangopadhyaya, A. Khare,
A. Pagnamenta and U. Sukhatme, Phys. Rev. \textbf{A48} 2786, (1993)

\bibitem{Sukhatme} U.P. Sukhatme, C. Rasinariu and A. Khare, Phys.
Lett. \textbf{A234} 401, (1997)

\bibitem{Carinena} J.F. Cari\~nena and
A. Ramos, J. Phys. \textbf{ A33} 3467, (2000); Rev. Math. Phys.
\textbf{12} 1279, (2000)

\bibitem{Faux} M. Faux and D. Spector, J. Phys. \textbf{A37} 10397,
(2004)

\bibitem{Sasaki} S. Odake and R. Sasaki, J. Math. Phys. \textbf{46}
063513, (2005); J. Nonlin. Math. Phys. \textbf{12} Suppl. \textbf{1}
507, (2005)

\bibitem{Deift2} P. A. Deift and E. Trubowitz, Commun. Pure Appl. Math.
\textbf{32} 121, (1979)

\bibitem{Gesztesy1} F. Gesztesy and R. Svirsky, Memoirs Amer. Math. Soc.
\textbf{118}(568) 1, (1995)

\bibitem{Gesztesy2} F. Gesztesy, B. Simon and G. Teschl, J. Analyse Math.
\textbf{70} 267, (1996)

\bibitem{Aleixo} A. N. F. Aleixo, A. B. Balantekin and M. A.
Candido-Ribeiro, J. Phys. \textbf{A36} 11641, (2003)

\bibitem{Combescure} M. Combescure, F. Gieres and M . Kibler, J. Phys.
\textbf{A37} 385, (2004)

\bibitem{Rudyak}
B. V. Rudyak and B. N. Zakhariev, Inverse Problems {\bf 3} 125,
(1987)
\bibitem{Schnitzer}
W. A. Schnitzer and H. Leeb, J. Phys. {\bf A26} 5145, (1993)

\bibitem{Cooper} F. Cooper, A. Khare, and U Sukhatme, Phys. Rep. \textbf{251} 267, (1995)

\bibitem{Junker} G. Junker, \textit{Supersymmetric Methods in Quantum and
Statistical Physics}, Springer, (1996)

\bibitem{Bagchi} B. K. Bagchi, \textit{Supersymmetry in Quantum and
Classical Mechanics}, Chapman, (2001)

\bibitem{Matveev} V. B. Matveev and M. A. Salle, \textit{Darboux
Transformation and Solitons}, Springer, (1991)

\bibitem{Cooper2001} F. Cooper, A. Khare, U. Sukhatme. \textit{Supersymmetry
in Quantum Mechanics}, World Scientific, (2001)

\bibitem{Cycon} H. L. Cycon, R. G. Froese, W. Kirsch and B. Simon, \textit{%
Schr\"odinger Operators-with Applications to Quantum Mechanics and Global
Geometry}, Springer, (1987)

\bibitem{Ince} E. L. Ince, \textit{Ordinary Differential Equations}, Dover
Publication, (1956)

\bibitem{Debergh} N. Debergh, Phys. Lett. \textbf{A219} 1, (1996)

\bibitem{Iachello} F. Iachello, Phys. Rev. Lett. \textbf{44} 772, (1982)

\bibitem{Nowakowski} M. Nowakowski and H. Rosu, Phys. Rev. \textbf{E65}
047602, (2002)

\bibitem{Adler}
V. E. Adler and A. B. Shabat, {\it Dressing chain for the acoustic spectral problem},
arXiv: nlin.SI/0604008

\bibitem{graham} R. Graham, Phys. Rev. Lett. \textbf{67} 1381, (1991)

\bibitem{socorro} J. Soccoro and E. R. Medina, Phys. Rev. \textbf{D61}
087702, (2000)

\bibitem{Balantekin} A. B. Balantekin, Phys. Rev. \textbf{D58} 013001, (1998)

\bibitem{A1}
A. A. Andrianov and F. Cannata, J. Phys. {\bf A37} 10297, (2004)

\bibitem{A2}
A. A. Andrianov, M. V. Ioffe, V.P. Spiridonov,  Phys. Lett. {\bf
A174} 273, (1993)

\bibitem{A3}
A. A. Andrianov, F. Cannata, M. Ioffe and D. Nishnianidze, Phys.
Lett. {\bf A266} 341, (2000)

\bibitem{A4}
M . V. Ioffe and D. N. Nishnianidze, Phys. Lett. {\bf A327} 425, (2004)

\bibitem{Natanzon}
G. A. Natanzon, Vestnik Leningrad Univ. {\bf 10} 22, (1971); {\it ibid} Teor. Mat. Fiz. {\bf 38} 146, (1979)

\bibitem{Ginocchio}
J. N. Ginocchio, Ann. Phys. (N.Y.) {\bf 152} 203, (1984), {\it ibid}
{\bf 159} 467, (1985)

\bibitem{Sing}
C. A. Sing and T. H. Devi, Phys. Lett. {\bf A171} 249, (1992)

\bibitem{Cooper2}
F. Cooper, J. N. Ginocchio and A. Khare, Phys. Rev. {\bf D36} 2458, (1987)

\end{thebibliography}
\end{document}